# An Empirical Study of How Users Adopt Famous Entities

Sheng Yu and Subhash Kak
Department of Computer Science
Oklahoma State University
Stillwater, U.S.A 74075
{yshe, subhashk}@cs.okstate.edu

***Abstract*— Users of social networking services construct their personal social networks by creating asymmetric and symmetric social links. Users usually follow friends and selected famous entities that include celebrities and news agencies. In this paper, we investigate how users follow famous entities. We statically and dynamically analyze data within a huge social networking service with a manually classified set of famous entities. The results show that the in-degree of famous entities does not fit to power-law distribution. Conversely, the maximum number of famous followees in one category for each user shows power-law property. To our best knowledge, there is no research work on this topic with human-chosen famous entity dataset in real life. These findings might be helpful in microblogging marketing and user classification.**

## I. Introduction

Social networking service (SNS) with its online sites and applications consists of three parts: users, social links, and interactive communications. In the past years, this kind of service has advanced greatly and changed our lives. Three worldwide popular SNS providers, Twitter, Facebook, and Tencent (qq.com), demonstrate the explosive growth and profound effect of this service. These three providers are in the top 10 most-visited websites in the world according to Alexa ranking [1]. For example, Tencent Weibo, one of the major products of Tencent Inc, has 425 million registered users and 67 million daily users.

Within the social networking service, users might mirror social relations in real life, build new social connections based upon interests and activities, or both. When building new social links, users typically adopt some famous entities. For example, on Twitter, a user might follow BBC Breaking News (@BBCBreaking) for news and Johnny Depp (@J0HNNYDepp) for personal preference.

In this paper, based upon a large-scale dataset with 1.94 million users and 50.66 million directed links in a real social networking service, we analyze how users follow famous entities with human-chosen category labels. The result confirms that there is power law phenomenon in users' adoption of famous entities. Additionally, we discuss why the power law applies to this situation. The result could be useful for improving microblogging marketing and user classification.

The rest of the paper is organized as follows: in section 2, some technical backgrounds and related research works are introduced. Section 3 describes the dataset. In sections 4 and 5, static and dynamic analyses of the dataset are provided. Discussion is provided in section 6 and conclusions are presented in section 7.

## II. Technical Background and Related Works

### *A. Social network*

A social network is the social structure with persons or organizations, which usually are represented as nodes, and social relations, which correspond to the connections among nodes. The social relation could be both explicit, such as kinship and classmates, and implicit, as in friendship and common interest. The small world and the scale free network are two classes of social networks with different structural relationships.

When a social network is viewed as a small world network, most nodes can reach every other node through a small number of links [1][2]. In real world, the famous theory of "six degree of separation" suggests that, on average every two persons could be linked within six hops. The situation in online SNS is somewhat different. The average distance on Facebook in 2008 was 5.28 hops, while in November 2011 it was 4.74 [3]. In MSN messenger network, which contains 180 million users, the median and the 90th percent degree of separation are 6 and 7.8 respectively [4]. And on Twitter, the median, average, and 90th percent distance between any two users are 4, 4.12 and 4.8 respectively [5]. In other words, the degree of separation varies on different SNS platforms and it changes with time.

Many properties of social networks show scale free phenomenon [6][7], that is, their degree distribution asymptotically follows a power law. For example, on Twitter, the number of followees/followers fits to the power-law distribution with the exponent of 2.276 [5]. In addition, the number being retweeted and mentioned by other users on Twitter also follows a power law [8].

1. http://www.alexa.com/topsites

*B. Social networking service*

Social networking service (SNS) embraces collections of online websites, applications, and platforms, which allow users to build social network and provide additional service [9][10]. The social network could be symmetric or asymmetric. In symmetric SNS, such as Facebook, the undirected social relations must be confirmed by both peers. Conversely, in an asymmetric SNS, such as Twitter, the directed social link could be made without the explicit permission from the destination user.

Different users publish their opinions and experiences via SNS, which aggregates personal wisdom and different standpoints. If extracted and analyzed properly, the data on SNS can lead to successful predictions of some human related events in near future [11][12].

In this paper, we focus on the structure of asymmetric SNS as microblogging service. First of all, we present some definitions.

*Follow*: user *A* follows user *B* means that there is a directed social link from *A* to *B*.

*Follower/Followee*: if user *A* follows user *B*, *B* is a followee of *A*, and *A* is a follower to *B*.

*Famous entity*: famous entities are specific users on social networking service that typically are celebrities, famous organizations, or some well-known groups. In this paper, we focus on the followees being famous entities, which are named as *famous followees*.

*C. Motivation to study famous followees*

On SNS, the user could be roughly classified into three categories: information source, information seeker, and friend [13]. Because a *nobody* is hard to be a qualified long-term information source, we safely assume most of the information sources are famous entities. Thus as information seeker, the user will follow different kinds of famous entities for different information. For example, for getting emerging news, a user will follow some news agencies and sports fans will follow sports superstars.

But how do users follow these famous entities? Do users follow lots of famous entities in one category, such as many computer experts? To our best knowledge, there is no research work that deals with human-chosen famous entity dataset in real life. The answer to these questions will be useful to understand users' activities and in further research with applications in microblogging marketing, as will be discussed in section 6.

## III. Datasets

In this section, we introduce the dataset, and give some basic characteristics of it.

Our dataset is published by Tencent Weibo for KDD Cup 2012[2]. Tencent Weibo was launched in April 2010, and is currently one of the largest microblogging providers in China. As a major platform of SNS, it has 425 million registered users, 67 million daily users, and 40 million new messages each day.

2. http://www.kddcup2012.org/c/kddcup2012-track1/data

The dataset is a sampled snapshot of Tencent Weibo, including user profiles, social graph, famous entities categories, famous followee adoption history, and so on. In this paper, we only use the three datasets, social graph, famous entities categories, and famous followee adoption history, for our analysis.

*Social graph*: contains all the following information at the sample time of the selected users, who were the most "active" ones during the sampling period.

*Famous entities categories*: include all the information of famous entities. A famous entity is a special user in Tencent Weibo to be recommended to other users. Typically, well-known celebrities, organizations and groups are selected to be the famous entities. All the famous entities and their categories are chosen and assigned by Tencent Inc.

*Famous followee adoption history*: indicates that records of users' new adoption of famous items in the sampling period. This dataset contains both rejections and acceptances records.

The second and third dataset will be discussed when they are used in section 4 and 5 respectively. And following is a brief description of the social graph dataset.

There are 1,944,589 users, including 1,892,059 followers, 920,110 followees in the dataset. Because this is a sampled snapshot, the dataset is asymmetrical. With 50,655,143 social link records for followers, average out-degree is 26.77, and for followees, average in-degree is 55.05. The distributions of the out-degree of followers and in-degree of followees are partly shown in Fig. 1 and Fig. 2, respectively. Similar to the results in previous researches [5][14], we find that both the out-degree and in-degree distributions fit to power-law.

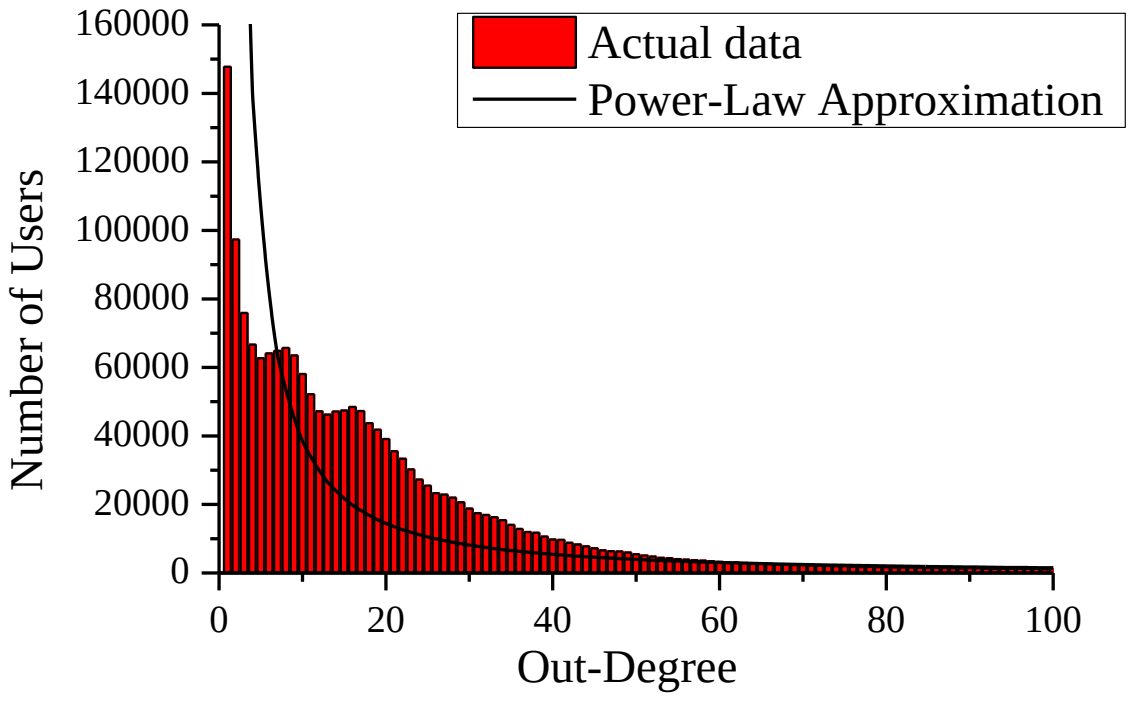

Fig. 1. The out-degree distribution of followers

Among 1,892,059 followers, there are 83,474 users following more than 100 followees. These 83,474 followers account for only 4.41% of the population, and are not included in Fig. 1. In total, the minimum, median, 90th percent, and maximum out-degree are 1, 14, 52, and 5188, respectively. Only considering the data in Fig. 1, the out-degree distribution approximately fits the following power-law distribution with R2 of 0.858:

$$Number_of_Users = 10^{6} \times Out_Degree^{-1.415} \quad (1)$$

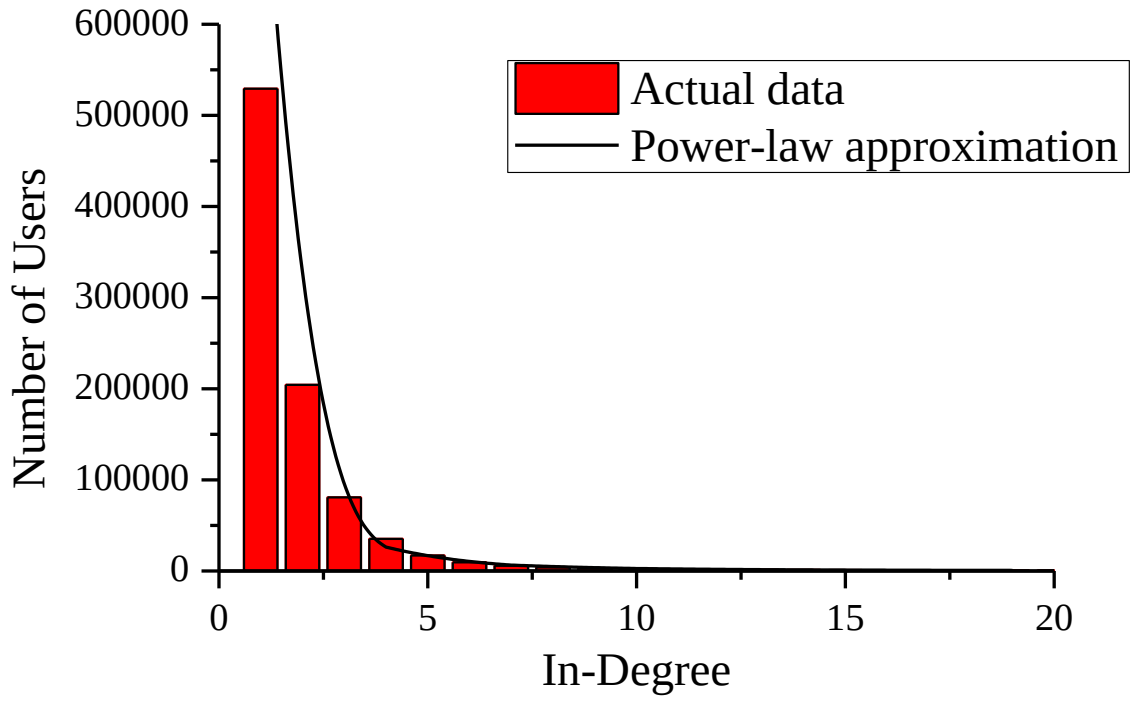

Fig. 2. The in-degree distribution of followees

Out of these 920,110 followees, 19,538 or about 2.12% in proportion, are followed by more than 20 other users. These 19,538 followees are also not shown in Fig. 2. Overall, the minimum, median, 90th percent, and maximum in-degree are 1, 1, 4, and 456,827, respectively. Additionally, only taking the in-degree equal to be equal to or less than 20 into consideration, the in-degree distribution can be approximately represented as the following power-law equation with R2 being 0.9899:

$$Number_of_Users = 840935 \times Out_Degree^{-2.501} \quad (2)$$

## IV. Static Analysis

In this section, we will analyze how users follow famous entities based upon the static snapshot of users' social graph and the famous entities dataset.

There are 6,095 famous entities in the dataset. All the famous entities and their categories were chosen manually by Tencent Inc. But only 5,796 of them, about 95.09%, are involved in the social graph dataset. The distribution of these famous entities' in-degree is shown in Fig. 3, including 4,930 famous entities with equal to or less than 10,000 followers.

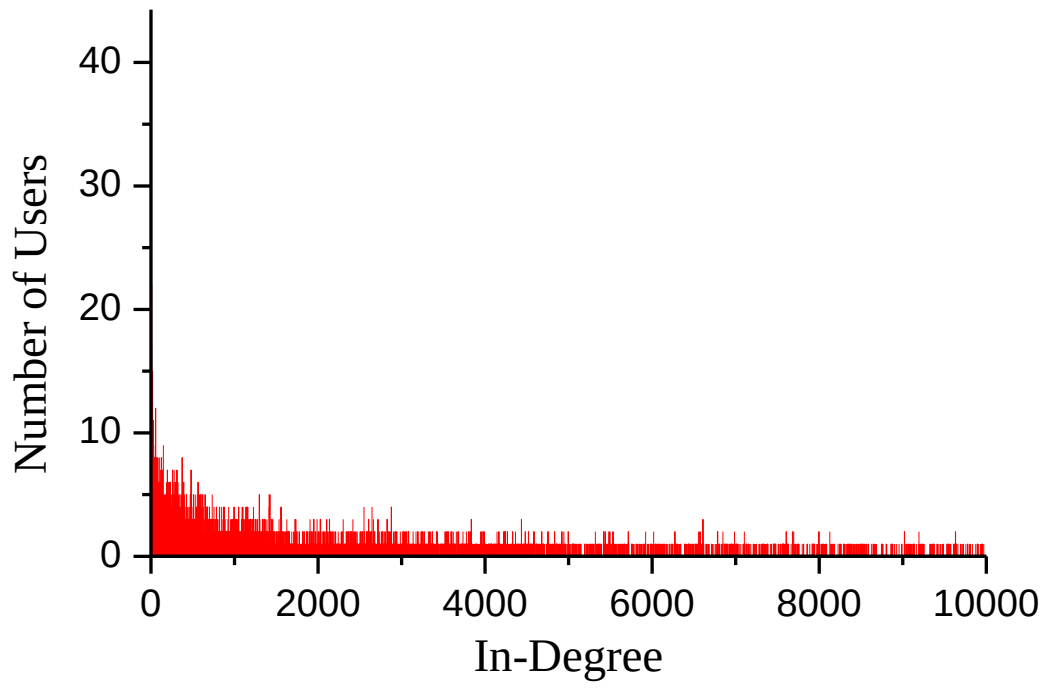

Fig. 3. The in-degree distribution of famous entities

There are 866 famous entities with more than 10,000 followers. These entities account for about 14.94% of the population, and are not shown in Fig. 3. Totally, there are 44,427,963 social links to these famous entities. Additionally, the minimum, median, average, 90th percent, and maximum in-degree are 1, 1,288, 7,665, 16,509, and 456,827, respectively.

Compared with the in-degree of overall users, which is shown in Fig. 2, the famous entities set has much more followers. Subjectively, these famous entities are well known to people and they are much more likely to be identified among millions of users. Additionally, and objectively on the Internet, the famous entities are more likely to be reliable and stable information sources. Consequently, the masses need to follow them to get needed information.

In Fig. 1 and 2, out-degree of followers roughly and in-degree of followees well fit to power-law distributions. Quite differently in Fig. 3, as a whole, the in-degree of famous entities is much more evenly distributed than the preceding two. The log-log plot of their in-degree is shown in Fig. 4 and there is no clear and strong linear correlation found in this figure. Thus it does not fit to power-law in any range.

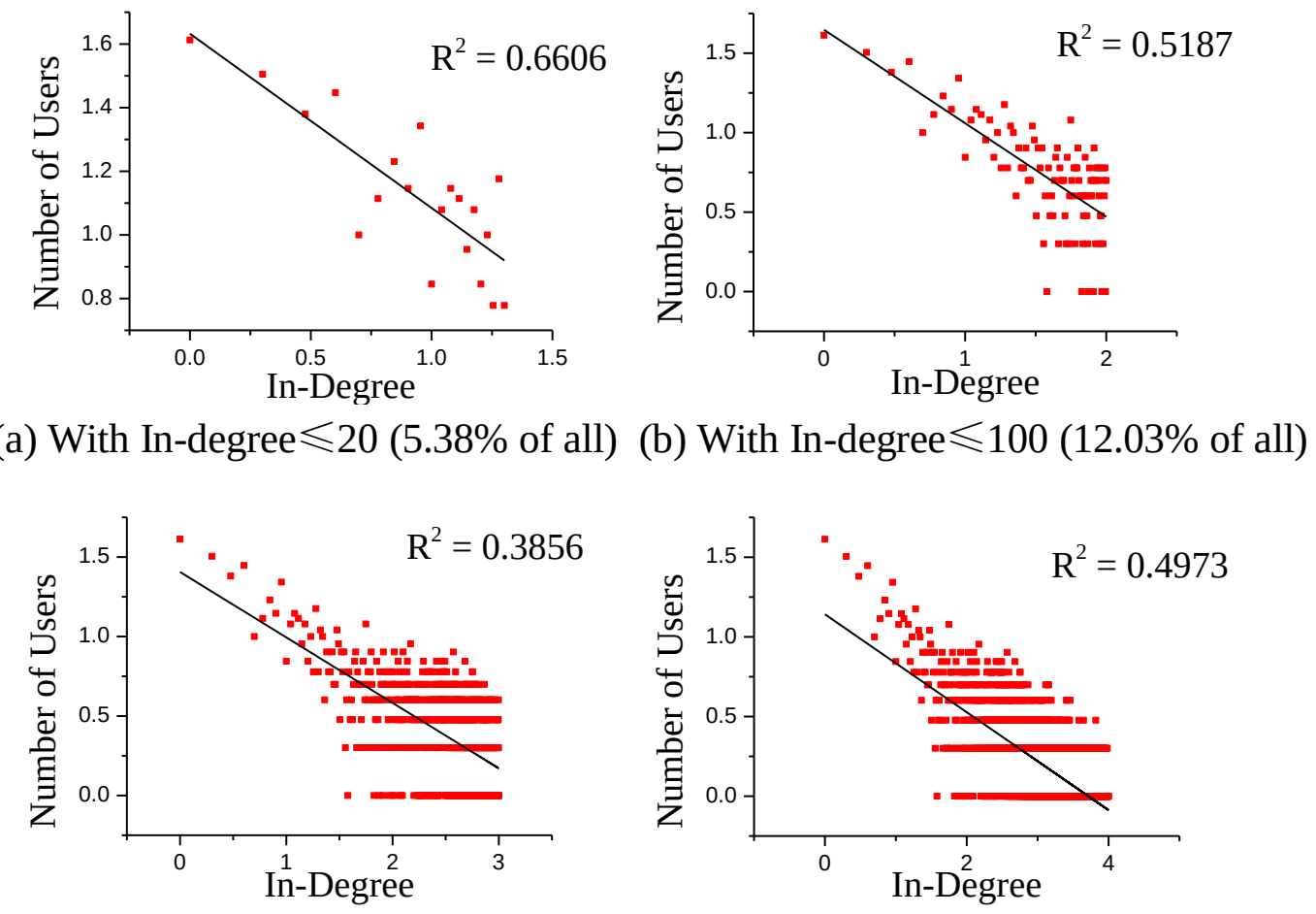

(a) With In-degree⩽20 (5.38% of all) (b) With In-degree⩽100 (12.03% of all)

(c) With In-degree⩽$10^3$ (44.74% of all)(d) With In-degree⩽$10^4$ (85.06% of all)

Red points: actual data Black lines: linear approximation Log base: 10

Fig. 4. The log-log plot of in-degree distribution of famous entities

Even though the famous entities generally have many followers, the number of followers of each one varies significantly. The mean value and standard deviation of in-degree for all famous entities are as high as 7,665 and 23,703 respectively. For these with in-degree ≤ 10,000, the mean value and standard deviation are 1,846 and 2,241 respectively. In other words, some famous entities may not get the same attentions on SNS as in reality. This also implies the importance of microblogging marketing. Without proper dissemination of information and marketing (that is, propaganda), it's hard to be a well-known user on SNS, even for a famous entity in real life.

Each famous entity has a hierarchical category, in form of "a.b.c.d". For example, for foursquare, one popular free application on mobile phones, the category could be: "science-and-technology.internet.mobile.location-based".

Within the famous entities dataset, there are 6 first-level, 27 second-level, 117 third-level, and 375 fourth-level categories. The famous entities are not uniformly distributed in each category. The number of famous entities in fourth-level categories is shown in Fig. 5. There are eight fourth-level categories with more than 100 famous entities in them. They have not been counted in Fig. 5.

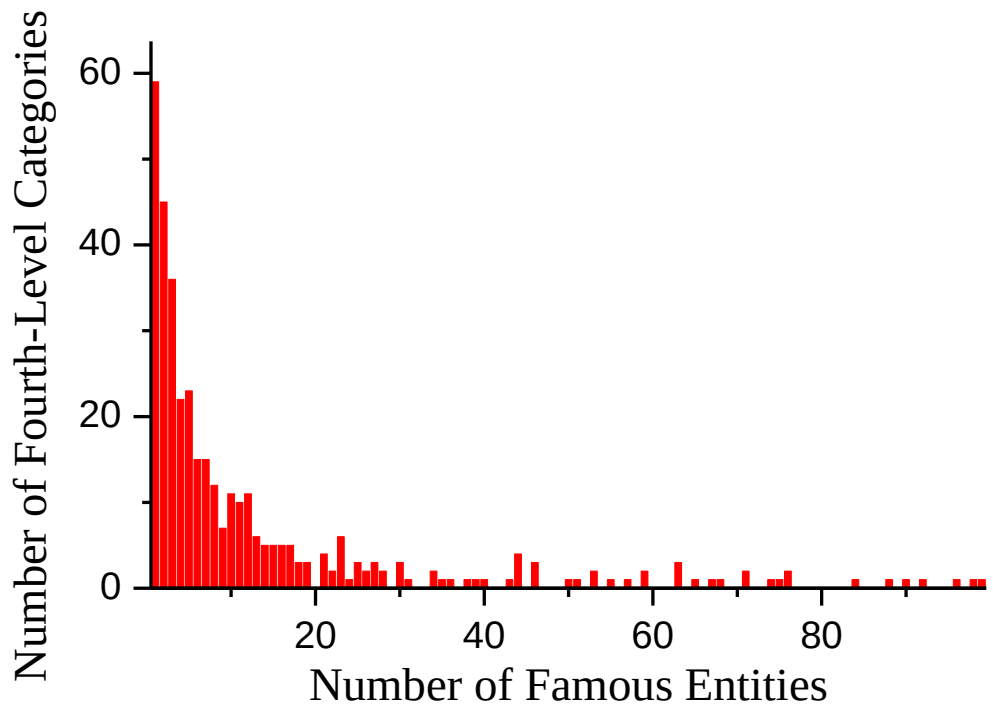

Fig. 5. The log-log plot of in-degree distribution of famous entities

In our analysis, we do not care about the hierarchical structure of the categories. Thus we could treat the dataset as 5,796 famous entities dispersing in 375 classes.

We measure the maximum number of famous followees in one category (*MFFC* for short in the following) for each user. After analyzing 50,655,143 social link records, only taking the users with famous followees into account, we obtained the distribution of maximum leaf value as shown in Fig. 6. There are 97,655 followers, about 5.16% of all, who do not have the social links to famous entities in the dataset, and hence are not involved in the following discussion.

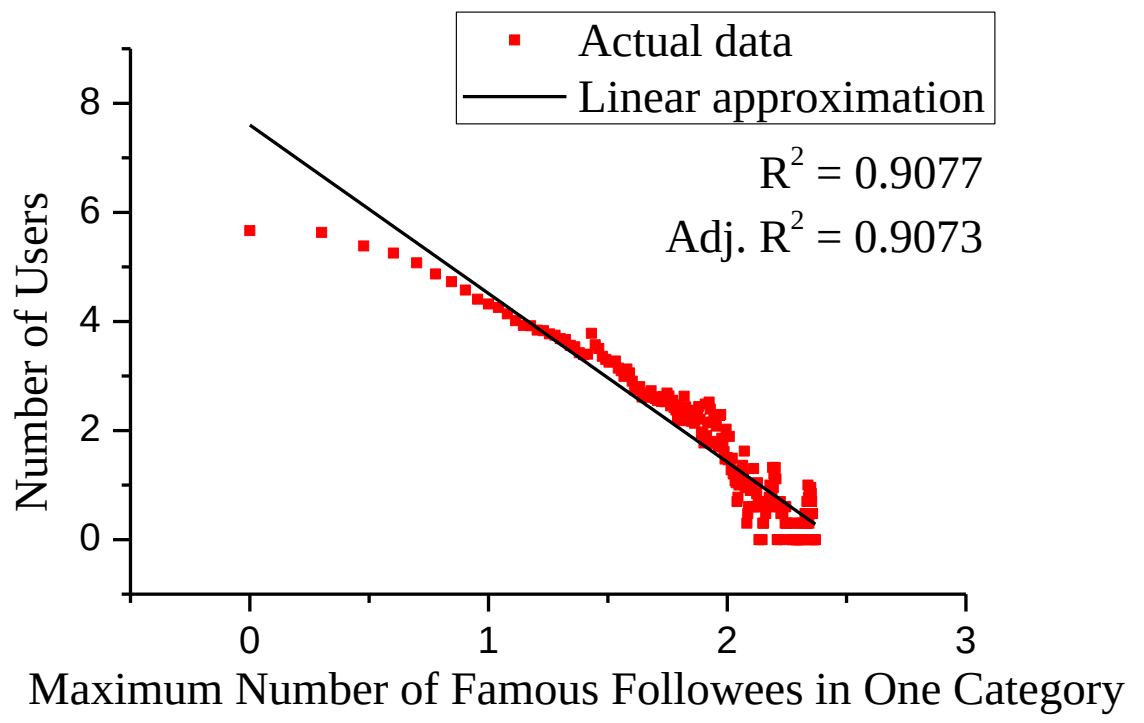

(a) In log-log plot in base 10

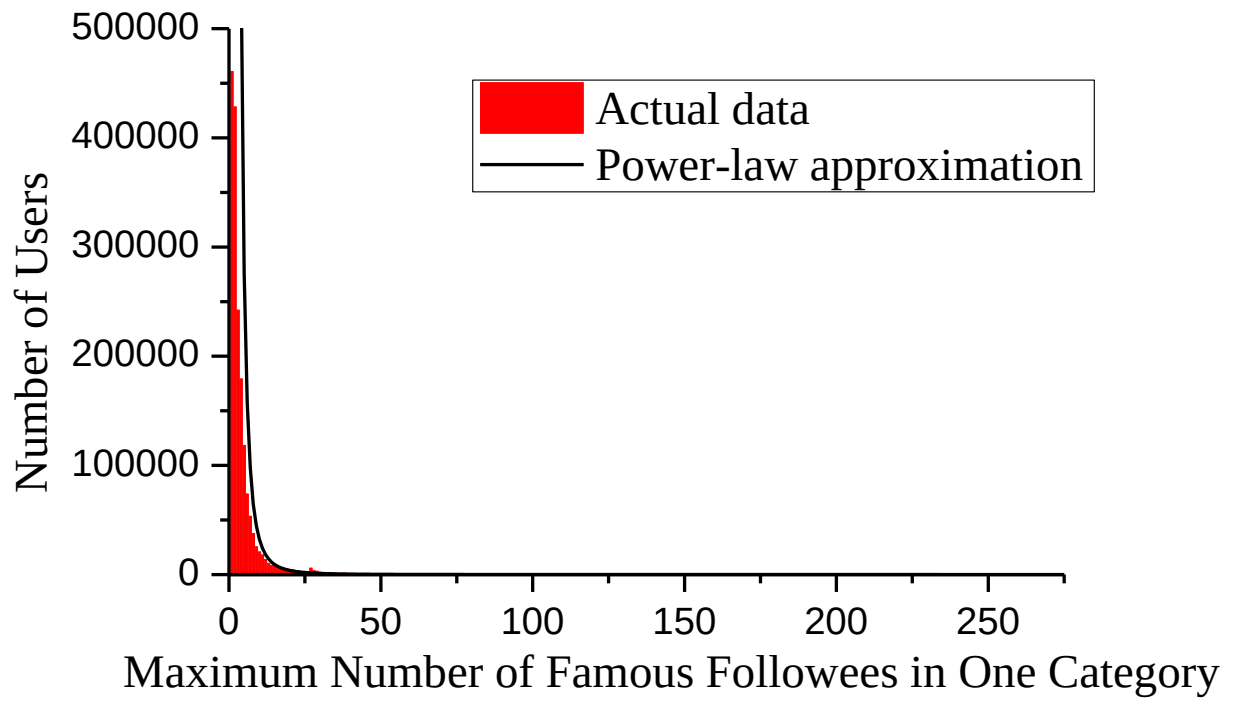

(b) In original coordinate

Fig. 6. Distribution of maximum number of famous followees in one category

Deriving from the linear correlation in Fig. 6(a), we get the power-law approximation as following in Fig. 6(b):

$$Numer_of_Users = 10^{7.601223} \times MFFC^{-3.08936} \tag{3}$$

Overall, with users of MFFC being 0, the minimum, median, 90th percent, and maximum "maximum leaf values" are 0, 2, 9, and 234, respectively. The average maximum leaf value is about 4.55. To make the power-law property clearer, we make the distribution in smaller rangers in Fig. 7.

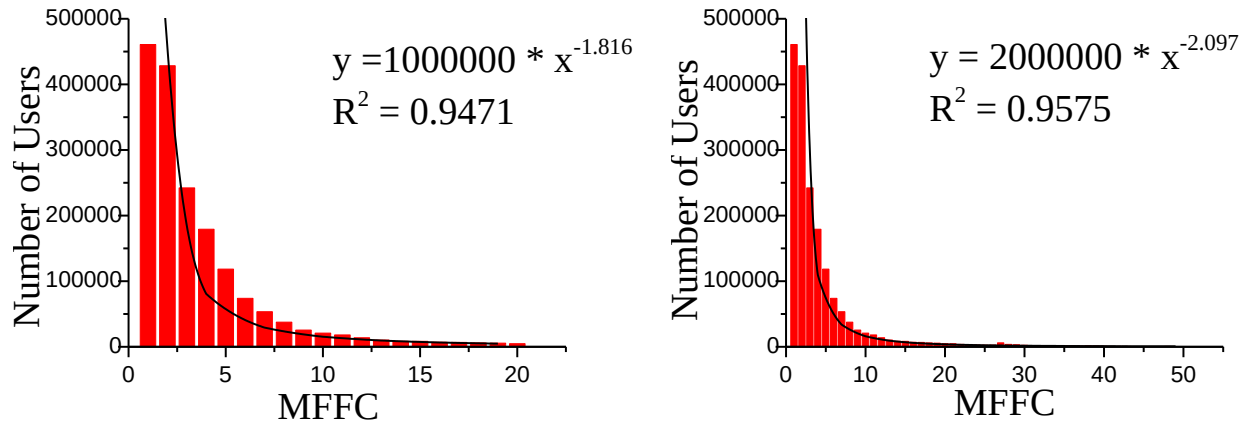

(a) With MFFC≤20 (96.39% of all) (b) With MFFC ≤50 (99.37% of all)

Red block: actual data Black lines: power-law approximation

Fig. 7. Distribution of the MFFC with part of users

Additionally, the percentage cumulative distribution is provided in Fig. 8.

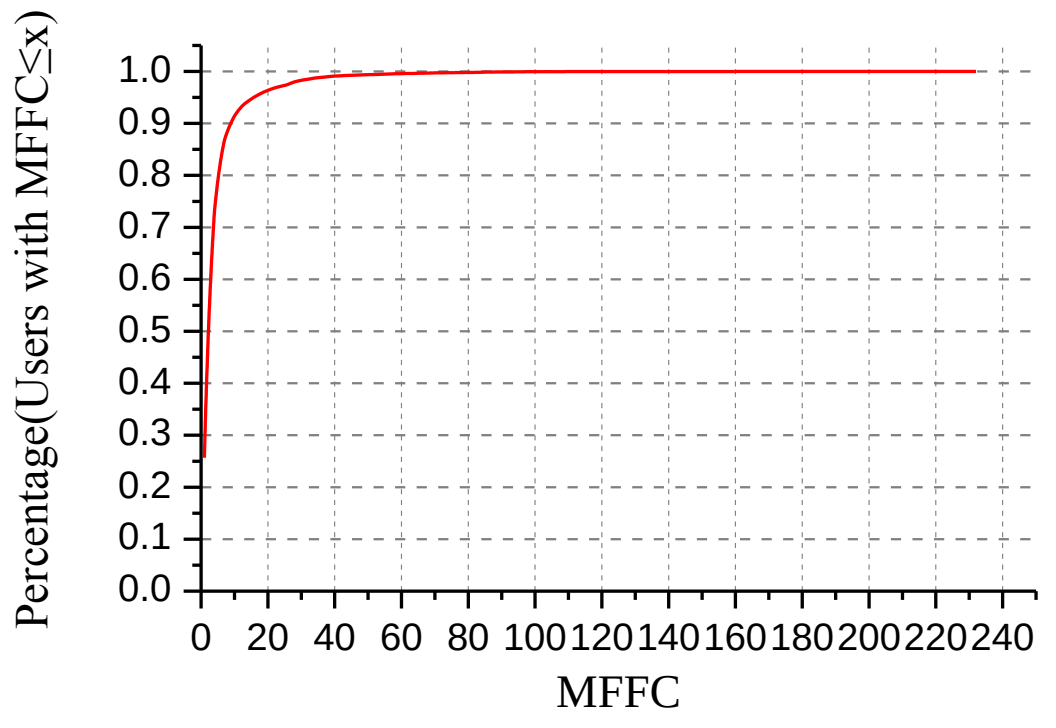

Fig. 8. Percentage cumulative distribution of MFFC

Generally, the users do no follower too many famous entities in one category and this phenomenon may be explained by marginal utility.

The users follow famous entities to seek information. At the beginning, the users could get lots of new information by starting following famous entities in one new category, because no alternative information source is available. After following a couple of famous entities in the interested realm, these information sources are enough to provide sufficient amount of messages. The gain from adoption of new users in that category has decreased and is little. If the users continue to adopt new famous entities in that field, information overload will become annoying and dissuade the user from adding new followees in the corresponding category. For instance, following 100 new agencies cannot give the user additional information, and only bother the individual with flood of duplicate messages.

With the long-tail property, a small fraction of users have lots of famous followees --- up to 234, in one category. Because all the information in these datasets is encoded as random strings or numbers to protect personal privacy and keep fairness in KDD Cup 2012, we cannot make deeper analysis of this matter in this paper. But we guess that the unusual and excessive adoption of famous entities in one single category may be related with the users' working and living environments. For example, an IT worker might follow more famous entities, in related categories of computer science, than others.

## V. Dynamic Analysis

This section provides a dynamic analysis using the users' adoption history and famous entities datasets.

The users' adoption history contains the users' choice, both rejections and acceptances, to the recommendations from Oct 11, 2011 to Nov 11, 2011. Totally there are 73,209,277 records in this dataset. But the following two kinds of records are removed and not used:

1. The follower in the record does not have its social links information in the social graph dataset;
2. The followee in the record is not a famous entity in our dataset.

Consequently, there are 62,169,578 (84.92% of all) valid records in this dataset. Because a user could accept a recommendation to follow one famous entity, then unfollow it, and accept the same recommendation again after some time. There are some repeated records with different timestamps and we did not remove them.

For each user, the adoption rate for a specific category is defined as following:

$$\text{Adoption Rate} = \frac{\text{Acceptances}}{\text{Rejections} + \text{Acceptances}}$$

The adoption rates for all users are shown in Fig. 9. According to Fig. 6, more than 90 percent of users have 9 or less famous followees in the maximum category. As a result, the samples of acceptance rates for the cases, in which the number of famous followees in one category is more than 9, are not sufficient relatively. Thus we combine all these cases into one class as "10+" in Fig. 9.

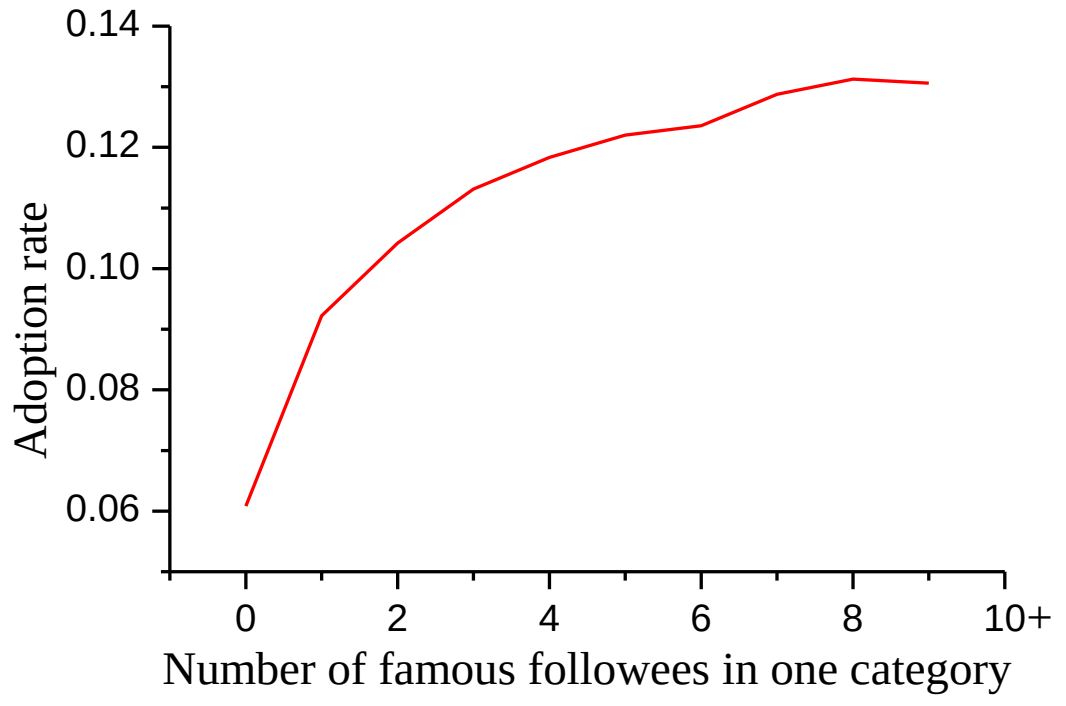

Fig. 9. The distribution of adoption rate

In the beginning, the adoption rate increases rapidly. But with the incensement of famous followees in one category, the adoption rate of that category grows much slower.

Assuming that there is no cost for adopting new followees, in the interested fields, the users might like to follow as many famous entities as there are, to get the most information. Thus the more famous entities are followed in a specific category, the more interest is developed in that realm, and thus it is more likely to adopt new followees in the same field.

But in real life, adopting new followees needs more energy and time to digest the additional messages. Thus there is cost associated with adoption.

Overall, for more than 90 percent of users, they have 9 or less famous followees in one category. That's, 9 or less information sources in one field are enough to provide sufficient messages with affordable cost. There are also less than 10% of users, who follow more than most of masses. For these users, the value of new messages is much higher than the cost. Thus even though users could get a little additional information by recruiting more followees, they continue to adopt new ones. For example, an analyzer of a company might follow all of its competitors, no matter how many they are.

In addition, the result in Fig. 9 fits with the general conclusions of section 4. To confirm this, we make an iterative simulation.

Initially, we set the number of users as 1,892,059, which is the same as the number of followers in the social graph dataset. And the MFFC of all users are set to 0 at the beginning.

In each iteration round, each user has one opportunity to increase its MFFC by one with the same possibility in Fig. 9. Because we are short of samples to evaluate the adoption rate for MFFC$\geq$10 well, the maximum MFFC of users in this simulation is limited to 10.

After 24 rounds, we could get the result, which is most similar with the real situation, as shown in Fig. 10. Except the MMFC of 0, the simulation matches the real situation very well. In one word, such adoption rates lead to the power law distribution in section 4.

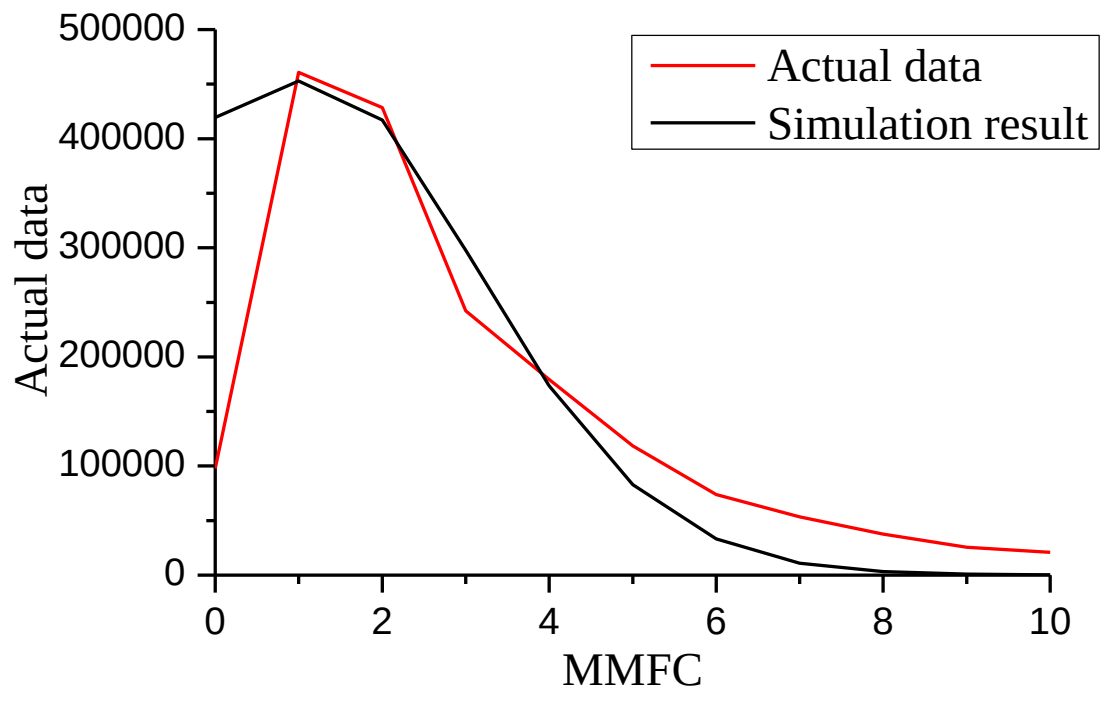

Fig. 10. The simulation result

For the case where MMFS is zero, the difference between real data and simulation could be an artifact of the sampling method. As we discussed in section 3, the sampled users are the most "active" ones, who are less likely not to adopt any famous entity, compared with the non-selected "inactive" ones. In real life, overall, the users with MMFC equal to 0, including the zombie accounts, should be represented by a greater proportion than is true of our samples datasets.

Theoretically, our case is similar with the classic Barabási–Albert model [15]:

1. Expand continuously: in terms of individuals, when they enter the SNS system, they commonly follow some other users in a short time, and then continue to adopt selective followees with a relatively slower pace. In terms of whole system, the existing users continue to choose new followees, and the new users continue to enter the system.
2. Rich get richer: with more famous followees, users are more likely to adopt more. But the increase rate in our case is different from the Barabási–Albert model.

In sum, both analyses confirm that, the MMFC of users fits to power law distribution.

## VI. Discussion

### A. Potential value

The results of this study could be useful in microblogging marketing. The microblogging marketing personnel could discover potential customers better with the results of this paper.

Consider that users with less than 3 followees in the corresponding category do not show significant interest in this field, and have a relative low adoption rate for recommendations. So if the microblogging marketing personnel propagate themselves to these users, the efficiency will be low, because of low adoption rate.

On the other hand, users following more than 9 entities in the category show greatest interest, and have the highest adoption rate. But the number of these users is small. And according to their extraordinary interest in the realm, they are not likely to be common users.

As a result, on balance, users who follow 3 to 9 famous entities in the category are the best ones to be targeted for promotion.

Furthermore, analyzing the distribution of users' followees will be helpful in automatic classification the users. If some users follow many more entities in a single category than most of the masses, they show an extraordinary interest in corresponding field. Such information could be used to find these "uncommon" users and classify them accordingly.

### B. Limitation

The used datasets were sampled and provided by Tencent Inc. and we chose the most active users in the sampling period. But the datasets do not provide the precise definition of "active" users. We do not know the standard by which the famous entities were chosen and labeled so by the employees of Tencent Inc. Finally, the recommendation algorithm will have some impact on the result in section 5. But at the statistical level, a few outliers cannot affect the general trend. A deeper examination of how different factors affect the results will be studied in the future.

### C. Future works

There are still open issues in this topic. For example, we provide confirmation that the maximum number of famous followees in one single category fits to power law. But the factors which affect the upper limit of famous followees in a category for each user are not clear and the model of adopting famous followees is not provided.

In addition, why the in-degree of famous entities varies greatly is unknown. It appears that being well-known in real life does not guarantee success on SNS. This question requires further research from the perspective of microblogging marketing.

## VII. Conclusion

Combining the static and dynamic analysis of datasets within a huge social networking service, we show how users follow famous entities. The results in both experiments show that the in-degree of famous entities does not fit to power-law distribution, while maximum number of famous followees in one single category does. These findings might be helpful in microblogging marketing and user classification.

## Acknowledgment

We thank Tencent Inc, the organizers of KDD Cup 2012, for sharing the datasets of microblogging service with the public.